# A Time Domain Coupled Electronic-Optical Simulation Model for GaN-based LEDs[I]

Mohammed Zia Ullah Khan[1,]

Department of Engineering Physics, Polytechnique Montreal, Quebec, Canada
Department of Electrical Engineering, King Fahd University of Petroleum & Minerals, Dhahran, Saudi Arabia

Mohammed. A. Alsunaidi[2,]

Department of Electrical Engineering, King Fahd University of Petroleum & Minerals, Dhahran, Saudi Arabia

Abstract

In this paper, we present a coupled carrier-photon model that accounts for the time-domain interactions between carrier transport and light emission in Gallium Nitride (GaN)-based LEDs that hasn't been reported so far. Carrier transport is modeled using the drift-Diffusion formulation, whereas light emission and propagation is modeled using Maxwell's equations.

The drift Diffusion equation is solved self-consistently with Poisson equation. The carrier transport and photon emission are coupled by formulating an appropriate relation between radiative recombinations and dipole sources such that the strength of the dipole sources is given by the radiative recombination rates gauged to an appropriate value. This gauging factor is obtained by calibrating the External Quantum Efficiency (EQE) of homojunction GaN LED with the experimental value.

[I]This document is a collaborative effort
Corresponding Author

Email addresses. mohammed-zia-ullah.khan@polymtl.ca (Mohammed Zia Ullah Khan), msunaidi@kfupm.edu.sa (M. Alsunaidi),
   URL. http.//faculty.kfupm.edu.sa/ee/msunaidi// (M. Alsunaidi),
http.//sa.linkedin.com/pub/mohammed-zia-ullah-khan/3a/9b6/682/ (Mohammed Zia Ullah Khan)
   [1]This is the first author
   [2]This is the second author

———

1. Introduction

Solid State Lighting (SSL) is expected to reduce electricity consumption for lighting by at least 50% [1]. This can be achieved through the development of light sources that consume less power or have high lumens per watt (lm/W) ratio. High power semiconductor light emitting diodes (LEDs) in the visible spectrum for SSL have been possible with the advent of direct bandgap III-Nitride semiconductor materials that have approached luminous efficiencies of 100 lm/W [2]. Group III-Nitride semiconductor materials such as Gallium Nitride (GaN), Indium Nitride (InN) and Aluminum Gallium Nitride (AlGaN) have been used to produce high brightness LEDs for SSL applications and are the subject of current research in SSL to improve device performance and to achieve high LED efficiencies [3, 4]. Due to the unique material properties of these semiconductors, they have been demonstrating high internal quantum efficiencies (IQE) compared to the commercial Silicon carbide (SiC) LEDs.

Future development and improvements in GaN LEDs require a deeper understanding of the physical nanoscale mechanisms and internal physics of the device. The parameters needed to improve the device performance can be investigated through numerical simulations of advanced models. The challenges that occur in modeling such devices are knowledge of a large number of III-Nitride material parameters and integration of carrier transport with optical domain. Numerical simulations help in developing the link between material properties and the performance of the device. They can also help to understand the nanoscale effects influencing the LED performance that cannot be explained experimentally. Therefore, advanced numerical simulation models of such LEDs have to be developed for future advancements. In this work, a coupled carrier-photon simulation model for GaN based LEDs is developed to study and analyze LED structures for SSL applications.

The development of numerical models is important for LED optimization and development and also to explain physical mechanisms through formulation of theoretical models that can explain these affects. Numerous simulation studies have been investigated for electronic, optical and thermal problems using commercial simulation softwares such as APSYS, COMSOL, and SiLENSE etc. These simulation programs are based on the drift diffusion (DD) transport model, polarization models for heterostructure LEDs and radiative and non-radiative recombination models. These programs use approximate ray tracing models to determine the optical characteristics such as photon extraction, emission spectrum and optical power. To analyze a practical LED structure, a full-wave simulation model consisting of both the carrier transport and an appropriate optical model coupled with the carrier transport have to be investigated. In this work, we focus on developing a complete optoelectronic model that simulates electronic and optical proper-ties of GaN LED structures in time domain and also incorporates the coupling between the optical and electronic parameters.

Thus, numerical simulations based on the DD transport model have been investigated for development and improved performance of these devices. 3D simulations of UV LEDs made from AlGaN/GaN quantum well structures were demonstrated for short range communication and SSL applications in [5]. A blue InGaN LED was studied using APSYS simulation program for optical performance and quantum efficiencies [6]. This LED had different compositions of Indium in the GaN-InGaN-GaN barrier layer and an optimum value of indium composition and number of barrier sub layers to obtain good optical performance. The studies performed using such simulation models focused on major electronic parameters i.e. recombination rates or internal quantum efficiency. Major issues that were addressed using these simulation studies were. The decrease in the internal quantum efficiencies (IQE) with the increase in the injection current termed as Efficiency droop [7], the effect of joule heating or temperature reducing the LEDs IQE [8], the effect of thickness of intrinsic GaN layers between multiple quantum wells (MQW) and p-GaN layer on light output powers [9], the effect of built-in spontaneous and piezoelectric polarizations on the luminescence that reduces the emission energies of active layers [10], electron leakage effects on IQE and radiative recombination rates in the active region that reduces these parameters [11].

Simulation studies focusing on improving the optical characteristics such as light extraction efficiencies in a GaN LED structure have also been addressed. Photonic crystal LEDs i.e. periodic arrangement of materials of different refractive indices on top of LEDs that can alter the propagation of photons such that total internal reflections at air/emitter interface are reduced have been demonstrated [12, 13]. Light extraction efficiencies were also improved by using metal gratings, surface roughening or texturing and these studies were demonstrated through FDTD simulations [13, 14]. 3D FDTD simulation studies of LED structures that can increase the light out-put power were investigated through patterning in sapphire substrates such that the light can be scattered or redirected inside the LED and escape from GaN layers [15].

Therefore, these studies focused on improving internal quantum efficiencies or extraction efficiencies either through electronic simulation or optical simulations but not both simultaneously. The major deficiencies observed in such simulation studies are following.

- The simulation models that link photon emission with the carrier trans-port describing the complete electronic and optical characteristics were missing.

- Appropriate relation between electron-hole pair recombinations and current density inputs to the optical model was absent.

- The dependence of light extraction efficiencies on electronic parameters was not demonstrated.

- The dependence of carrier mobilities and Diffusion coefficients on applied bias was not classified.

- The effect of hole transport parameters on GaN LED performance was missing.

Therefore, in this work we attempt to develop a coupled electronic-optical simulation model for GaN LEDs. To achieve this task, a time domain electronic model should be formulated in order to perform simulation studies in conjunction with optical model using FDTD numerical technique.

2. Formulation of the Coupled Model Algorithm

The coupled electronic-optical numerical model is developed based by linking the carrier transport with the photon emission. The radiative recombination rate obtained from the solution of carrier transport equations is coupled with the electric and magnetic fields obtained from the time domain solution of Maxwell's equations.

The electronic model describing carrier transport is given by the following set of coupled equations (1)-(3) i.e. the Poisson's equation, the current density equation and the carrier continuity equations [16, 17]

$$\nabla .\left( \varepsilon \nabla V \right) = -\rho \qquad (1)$$

$$\mathbf{J_n} = -qn\mu_n \nabla V + qD_n \nabla n, \ \mathbf{J_p} = -qp\mu_p \nabla V - qD_p \nabla p \qquad (2)$$

$$\frac{\partial n}{\partial t} = G_n - R_n + \frac{1}{q}\nabla .\mathbf{J_n}, \ \frac{\partial p}{\partial t} = G_p - R_p - \frac{1}{q}\nabla .\mathbf{J_p} \qquad (3)$$

where $V$ is the electrostatic potential, $\varepsilon$ is dielectric permittivity, $\rho$ is the volume charge density, $J_n$ and $J_p$ are electron and hole current densities, n and p are electron and hole densities, q electronic charge magnitude, $\mu_{n(p)}$ and $D_{n(p)}$ are carrier mobility and diffusion coefficient, $G_{n(p)}$ and $R_{n(p)}$ are electron-hole pair generation and recombination rates, respectively. $\nabla .(.)$ is the divergence operator and $\nabla (.)$ is the gradient operator.

On the other hand, the optical model describing the photon or light propagation and emission in GaN LEDs is obtained from the solution of the following set of Maxwell's equations (4)-(6)

$$\nabla \times \mathbf{H} = \mathbf{J}_{ext} + \frac{\partial \mathbf{D}}{\partial t} \qquad (4)$$

$$\frac{\partial \mathbf{B}}{\partial t} = -\nabla \times \mathbf{E} \qquad (5)$$

$$\nabla .\mathbf{D} = \rho_{ext}, \ \nabla .\mathbf{B} = 0 \qquad (6)$$

where D, E, B, H and $J_{ext}$ are the electric flux density, electric field intensity, magnetic flux density, magnetic field intensity and current density vectors respectively. The photon emission is related to the band to band or radiative recombinations. Its propagation in an LED structure also depends on the absorption coefficient which is related to the absorbed power. Thus, there are two major terms that are involved in coupling the two models. First, the electron-hole pair recombinations for light generation. Second, the light absorption for the electron-hole pair generation. The first coupling parameter requires modeling an expression that represents electron-hole pair radiative recombination rate in terms of current density sources. Therefore, for a time domain simulation, the current dipole source $J_{ext}$ as function of radiative recombination rate $R_{rad}$ and a time source s(t) for a GaN LED can be written as

$$\mathbf{J}_{ext} \propto R_{rad} \times s(t) \qquad (7)$$

or,

$$\mathbf{J}_{ext} = K \times \mathbf{P}_r \times R_{rad} \times s(t) \qquad (8)$$

where $K$ is the proportionality constant or gauging factor in C-m, $P_r$ is a random vector for dipole sources for random polarizations such that $|P_r|^2 = P_x^2 + P_y^2 = 1$ and $0 < P_x 1, 0 < P_y 1$. The value of $K$ is calibrated in such a way that the EQE of the p-n homojunction GaN LED from simulation model matches with the experimental value [18]. s(t) is the continuous time signal with multiple output frequencies representing the continuous light emission through radiative recombinations. The modeling of time source s(t) is discussed in the subsequent section.

The second parameter requires to link the optical carrier generation rate (Eq. 9) calculated from the EM fields to the carrier continuity equation.

$$G = \frac{Absorbed\ power}{Photon\ energy} = \frac{W}{E_{ph}} \qquad (9)$$

where, $W = -\nabla \cdot \vec{S}$, $\vec{S} = \vec{E} \times \vec{H}$, $\vec{S}$ is the power density and $E_{ph} = \frac{hc_0}{\lambda}$ is the photon energy.

3. Numerical Solution of the Coupled Model

The time domain simulation of GaN LEDs is performed using Finite difference time domain (FDTD) numerical method which is a popular technique that has been widely used in computational electromagnetics problems. Several approaches have been made to numerically solve the carrier transport equations such as Monte-Carlo methods, Finite-Difference (FD) and Finite-Element (FE), Quasi Fermi level models and Quantum transport models [16, 17, 19, 20]. However, these models involve steady state analysis that solves the carrier concentrations from the Fermi levels instead of time dependent continuity equations. Similar approaches were made to solve Maxwell's equations using numerical methods such as Beam Propagation Method (BPM), Method of Lines (MoL), Finite- Difference Time Domain (FDTD), Finite Element Method (FEM), Method of Moments etc. [21].To analyze the time domain carrier-wave interactions in a GaN LED, a time domain numerical method has to be used to solve both carrier transport and Maxwell's equations. Therefore, we propose a time domain coupled model to study electronic and optical characteristics inside GaN LED structures. The active device model is based on FDTD method and is coupled with time varying electric and magnetic fields.

Therefore, the FDTD algorithm for electronic-optical simulation of GaN LEDs is developed to simulate the carrier transport and photon emission. The resulting FDTD algorithm is used to model the active part of the device by solving the drift diffusion (DD) transport equations (1)-(3) and the electromagnetic part by solving Maxwell's equations (4)-(6) in time domain.

3.1. Electronic model FDTD algorithm

The closed form solutions of the basic carrier transport equations (1)-(3) for any semiconductor device are limited in applications and accuracy for nonlinear or multi-dimensional problems. Therefore, numerical techniques can be used to investigate such problems. This work uses the finite difference numerical technique to solve these set of coupled nonlinear partial differential equations. FDTD solution of carrier transport model produces the solutions for the electrostatic potential and the electron and hole densities (V, n and p). The partial derivatives are discretized by finite difference approximations. The Poisson's equation is solved for electrostatic potential using successive-over relaxation (SOR) iterative method. The Poisson's equation in (1) is solved in an iterative loop given by equations (16)-(19)

$$V^m(i,j) = V^{m-1}(i,j) + \omega R^{m-1}(i,j) \qquad (10)$$

where $\omega = 2\left[\cos\left(\dfrac{\pi}{i_e}\right) + \cos\left(\dfrac{\pi}{j_e}\right)\right]$, $and\ 0 < \omega < 2$, and $i_e$, $j_e$ are the number of grid points in $x$ and $y$ directions, respectively.

$$a(i,j)V^k(i+1,j) + b(i,j)V^k(i-1,j) + c(i,j)V^k(i,j+1) \qquad (11)$$
$$+ d(i,j)V^k(i,j-1) + e(i,j)V^k(i,j) = f(i,j)$$

$$R(i,j) = \dfrac{\begin{bmatrix} a(i,j)V(i+1,j) + b(i,j)V(i-1,j) + c(i,j)V(i,j+1) \\ + d(i,j)V(i,j-1) + e(i,j)V(i,j) - f(i,j) \end{bmatrix}}{e(i,j)} \qquad (12)$$

$$a(i,j) = \varepsilon(i,j), b(i,j) = \varepsilon(i-1,j),$$
$$c(i,j) = r\varepsilon(i,j), d(i,j) = r\varepsilon(i,j-1),$$
$$e(i,j) = (-1-r)\varepsilon(i,j) - \varepsilon(i-1,j) - r\varepsilon(i,j-1),$$
$$f(i,j) = q\Delta x^2 \left(n - N_D^+ - p + N_A^-\right), r = \frac{\Delta x^2}{\Delta y^2}$$
(13)

Similarly, the finite difference approximations of current density equations, electron and hole continuity equations (2)-(3) for a 2D simulation domain are obtained as

$$J_{n,x}^{k+\frac{1}{2}}\left(i+\frac{1}{2},j\right) = -q\left\{\frac{\mu_n(i+1,j) + \mu_n(i,j)}{2}\right\}$$
$$\left\{\frac{n^k(i+1,j) + n^k(i,j)}{2}\right\}\left\{\frac{V^k(i+1,j) - V^k(i,j)}{\Delta x}\right\}$$
$$+q\left\{\frac{D_n(i+1,j) + D_n(i,j)}{2}\right\}\left\{\frac{n^k(i+1,j) - n^k(i,j)}{\Delta x}\right\}$$
(14)

$$J_{n,y}^{k+\frac{1}{2}}\left(i,j+\frac{1}{2}\right) = -q\left\{\frac{\mu_n(i,j+1) + \mu_n(i,j)}{2}\right\}$$
$$\left\{\frac{n^k(i,j+1) + n^k(i,j)}{2}\right\}\left\{\frac{V^k(i,j+1) - V^k(i,j)}{\Delta y}\right\}$$
$$+q\left\{\frac{D_n(i,j+1) + D_n(i,j)}{2}\right\}\left\{\frac{n^k(i,j+1) - n^k(i,j)}{\Delta y}\right\}$$
(15)

$$J_{p,x}^{k+\frac{1}{2}}\left(i+\frac{1}{2},j\right) = -q\left\{\frac{\mu_p(i+1,j) + \mu_p(i,j)}{2}\right\}$$
$$\left\{\frac{p^k(i+1,j) + p^k(i,j)}{2}\right\}\left\{\frac{V^k(i+1,j) - V^k(i,j)}{\Delta x}\right\}$$
$$-q\left\{\frac{D_p(i+1,j) + D_p(i,j)}{2}\right\}\left\{\frac{p^k(i+1,j) - p^k(i,j)}{\Delta x}\right\}$$
(16)

$$J_{p,y}^{k+\frac{1}{2}}\left(i,j+\frac{1}{2}\right) = -q\left\{\frac{\mu_p(i,j+1) + \mu_p(i,j)}{2}\right\}$$
$$\left\{\frac{p^k(i,j+1) + p^k(i,j)}{2}\right\}\left\{\frac{V^k(i,j+1) - V^k(i,j)}{\Delta y}\right\}$$
$$-q\left\{\frac{D_p(i,j+1) + D_p(i,j)}{2}\right\}\left\{\frac{p^k(i,j+1) - p^k(i,j)}{\Delta y}\right\}$$
(17)

$$n^{k+1}(i,j) = n^k(i,j) + \frac{\Delta t}{q} \left\{ \frac{J_{n,x}^{k+\frac{1}{2}}\left(i+\frac{1}{2},j\right) - J_{n,x}^{k+\frac{1}{2}}\left(i-\frac{1}{2},j\right)}{\Delta x} \right\} \quad (18)$$

$$+ \frac{\Delta t}{q} \left\{ \frac{J_{n,y}^{k+\frac{1}{2}}\left(i,j+\frac{1}{2}\right) - J_{n,y}^{k+\frac{1}{2}}\left(i,j-\frac{1}{2}\right)}{\Delta y} \right\} + \Delta t \left( G_n^k(i,j) - R_n^k(i,j) \right)$$

$$p^{k+1}(i,j) = p^k(i,j) - \frac{\Delta t}{q} \left\{ \frac{J_{p,x}^{k+\frac{1}{2}}\left(i+\frac{1}{2},j\right) - J_{p,x}^{k+\frac{1}{2}}\left(i-\frac{1}{2},j\right)}{\Delta x} \right\} \quad (19)$$

$$- \frac{\Delta t}{q} \left\{ \frac{J_{p,y}^{k+\frac{1}{2}}\left(i,j+\frac{1}{2}\right) - J_{p,y}^{k+\frac{1}{2}}\left(i,j-\frac{1}{2}\right)}{\Delta y} \right\} + \Delta t \left( G_p^k(i,j) - R_p^k(i,j) \right)$$

The scalar terms n, p, $\mu_n, \mu_p$, $D_n$ and $D_p$ are averaged for consistency. The field dependence of transport parameters i.e. mobilities and diffusion coefficients is incorporated in the electronic model through appropriate models as discussed in [22, 23, 24, 25]. The Monte Carlo simulation data for the electron diffusion coefficient in GaN presented in [25] is incorporated in the electronic model through the curve t as shown in figure 3.

Therefore, at each time step, the new values of J are calculated using the previous values of V, n and p and are stored for next time step. Similarly, the new values of n and p are calculated using the previous values of n, p, and G and the new values of J. Thus, the update equations (20)-(25) develop the basic FDTD algorithm for studying electronic carrier transport in any semiconductor device. These basic semiconductor equations are solved with suitable boundary conditions i.e. Dirchilet and Neumann boundary conditions [16] for electrostatic potential and carrier densities. Dirchilet boundary condition is used to assign the thermal equilibrium values for potential and concentration at the metal contacts, whereas the Neumann boundary condition is used for continuity of potential and concentration at the non-contact or free surfaces.

### 3.2. Optical model FDTD algorithm

As discussed, the electromagnetic light wave propagation in GaN LEDs is described by the solution of the coupled partial differential equations representing the four Maxwell's equations (4)-(6). In this work, we use generalized Auxiliary Differential Equation (ADE)-FDTD algorithm to solve the electromagnetic part of GaN LEDs that also incorporates the optical properties of dispersive materials [26].

Equations (4)-(6) for a linear, non-dispersive isotropic medium can be written as

$$\frac{\partial \mathbf{H}}{\partial t} = -\frac{1}{\mu} \nabla \times \mathbf{E} \tag{20}$$

$$\frac{\partial \mathbf{E}}{\partial t} = \frac{1}{\varepsilon} \nabla \times \mathbf{H} - \mathbf{J} \tag{21}$$

Expressing equations (26)-(27) in scalar form, simplifying them for a 2D simulation in x-y domain and considering TM polarization because of the polar property of III-Nitrides, we obtain following set of equations

$$\frac{\partial E_x}{\partial t} = \frac{1}{\varepsilon} \frac{\partial H_z}{\partial y} - J_x \tag{22}$$

$$\frac{\partial E_y}{\partial t} = -\frac{1}{\varepsilon} \frac{\partial H_z}{\partial x} - J_y \tag{23}$$

$$\frac{\partial H_z}{\partial t} = \frac{1}{\mu} \left( \frac{\partial E_x}{\partial y} - \frac{\partial E_y}{\partial x} \right) \tag{24}$$

The central difference approximations of spatial and temporal derivatives in equations (26)-(27) utilizing Yee's spatial gridding scheme and a leap frog scheme gives the following update equations for solving E and H fields

$$\frac{E_x^{k+1}\left(i+\frac{1}{2},j\right) - E_x^k\left(i+\frac{1}{2},j\right)}{\Delta t} = \left( -\frac{1}{\varepsilon} \frac{\left[ H_z^{k+\frac{1}{2}}\left(i+\frac{1}{2},j+\frac{1}{2}\right) - H_z^{k+\frac{1}{2}}\left(i+\frac{1}{2},j-\frac{1}{2}\right) \right]}{\Delta y} - \frac{1}{\varepsilon}\left( J_x^{k+\frac{1}{2}}\left(i+\frac{1}{2},j\right) \right) \right) \tag{25}$$

$$\frac{E_y^{k+1}\left(i,j+\frac{1}{2}\right) - E_y^k\left(i,j+\frac{1}{2}\right)}{\Delta t} = \left( -\frac{1}{\varepsilon} \frac{\left[ H_z^{k+\frac{1}{2}}\left(i+\frac{1}{2},j+\frac{1}{2}\right) - H_z^{k+\frac{1}{2}}\left(i-\frac{1}{2},j+\frac{1}{2}\right) \right]}{\Delta x} - \frac{1}{\varepsilon}\left( J_y^{k+\frac{1}{2}}\left(i,j+\frac{1}{2}\right) \right) \right) \tag{26}$$

$$\frac{\left[\begin{array}{c} H_z^{k+\frac{1}{2}}\left(i+\frac{1}{2},j+\frac{1}{2}\right) \\ -H_z^{k-\frac{1}{2}}\left(i+\frac{1}{2},j+\frac{1}{2}\right) \end{array}\right]}{\Delta t} = \frac{1}{\mu}\left(\frac{\left[\begin{array}{c} E_x^k\left(i+\frac{1}{2},j+1\right) \\ -E_x^k\left(i+\frac{1}{2},j\right) \end{array}\right]}{\Delta y} - \frac{\left[\begin{array}{c} E_y^k\left(i+1,j+\frac{1}{2}\right) \\ -E_y^k\left(i,j+\frac{1}{2}\right) \end{array}\right]}{\Delta x}\right) \qquad (27)$$

$\Delta t$ (for two dimension case) for optical model is given by the Courant-Friedrich-Levy (CFL) stability condition [27]

$$(\Delta t)_{optical} \leq (\Delta t)_{max} = \frac{1}{c\sqrt{\left(\frac{1}{\Delta x^2} + \frac{1}{\Delta y^2}\right)}} \qquad (28)$$

where c is the velocity of light in the medium.

The optical properties of dispersive materials are modeled by using a general auxiliary differential equation (ADE)-FDTD algorithm that can include multi-pole Lorentz-Drude or Lorentz model [28]. For the case of materials that are modeled using multi pole Lorentz-Drude model, the update equation for polarization and electric field becomes

$$P_i^{k+1} = C_{1i}P_i^k + C_{2i}P_i^{k-1} + C_{3i}E_i^k \qquad (29)$$

$$E^{k+1} = \frac{\left(D^{k+1} - \sum_i^N P_i^{k+1}\right)}{\varepsilon_0 \varepsilon_\infty} \qquad (30)$$

with $C_{1i} = \frac{4d_i - 2b_i \Delta t^2}{2d_i + c_i \Delta t}, C_{2i} = \frac{-2d_i + c_i \Delta t}{2d_i + c_i \Delta t}, C_{3i} = \frac{2a_i \Delta t^2}{2d_i + c_i \Delta t}$

3.3. Coupled model FDTD algorithm

As discussed in section 2, the two major terms that couple the two models are the recombination rates for light emission and the generation rates for light absorption. The time domain expression for current dipole sources in terms of radiative recombination rates given by (8) can therefore be obtained as.

$$J_x(i,j)^k = K \times P_x(i,j) \times R_{rad}^k \times s(t) \qquad (31)$$
$$J_y(i,j)^k = K \times P_y(i,j) \times R_{rad}^k \times s(t)$$

This implies the current density source inputs are computed from radiative recombination rate obtained from the electronic model.

The time domain expression for optical generation rate given by (9) is obtained by discretizing (38).

$$G(x, y) = -\frac{\nabla \cdot S(x, y)}{E_{ph}} = -\frac{\left(\frac{\partial}{\partial x}(E_y H_z) - \frac{\partial}{\partial y}(E_x H_z)\right)}{E_{ph}} \quad (32)$$

Therefore the time domain optical generation rate generation rate computed from EM fields obtained from the solution of optical model is used for the computation of carrier concentrations using the electronic model carrier continuity equations. The coupling between the two models through generation and recombination rates is demonstrated in coupled model algorithm shown in figure 4.

4. Validation of Coupled Model Algorithm

The formulated model is validated in two steps. First, by verifying the electronic transport model by simulating a GaN p-n junction in 1-D. Next, the coupled electronic-optical model is verified by simulating a p-n homo-junction GaN UV-LED and comparing it with the experimental results.

4.1. Simulation of GaN p-n junction in 1D

The GaN p-n junction in 1D is simulated using electronic model FDTD algorithm at zero applied potential with the parameters given in Table 1. The thickness of p and n layers considered in simulation is 1 m with spatial and temporal step sizes of 10 nm and 0.5 fs respectively. The analytical and simulated values of built-in potential, depletion widths and volume charge densities are then compared as shown in figures 5 and 6. The simulated parameters are in good agreement with the calculated analytical values using expressions in [16].

4.2. Simulation of 2D p-n Homojunction GaN LED

A basic 2D p-GaN/ n- GaN homojunction UV LED with an emission wavelength at 365 nm is simulated. First, the structure is analyzed using the electronic model FDTD algorithm followed by optical simulation and then coupled model simulation. The simulated structure shown in figure 7 consists of a 300 nm p-GaN layer with a doping of $10^{17}$cm$^3$, 1 m n-GaN layer doped at $10^{17}$cm$^3$, 100 nm intrinsic GaN buffer layer and a 1 m sapphire substrate layer. The LED structure is biased to $V_p = V_{applied}$ $V_{bi}$ (p-contact) and $V_n = 0$ (n-contact). The FDTD simulation domain is 100 x341 grid points with x = y = 10nm and $\Delta t$ = 0.5f s. The simulations are run until the solutions of the electronic model reaches steady state.

The simulated steady state distributions of major parameters such as electrostatic potential, recombination rates and I-V characteristics are shown in figures 8-10. It can be seen from figure 8 that the potential starts from a high value at the p-contact and reaches low values as we go towards the side n-contact at forward bias conditions. It is evident from figure 10 that the radiative recombination rates are stronger than the non-radiative recombinations for the GaN recombination coefficients. It is also clear that the recombination rates are high at the junction resulting from the large number of electron-hole pairs at the junction as shown in figure 9.

The optical model or coupled model simulation requires modeling of time source explicitly. The time source in the case of GaN LEDs is a continuous wave (CW) dipole sources representing CW emission in the active layer as discussed in section 2. For the optical model, a Gaussian pulse line source with a carrier wavelength at GaN emission is modeled using equation 39.

$$J_x(t) = J_y(t) = A exp\left(-\left(\frac{t-t_0}{t_p}\right)^2\right) \cos\left(\frac{2\pi}{\lambda}t\right) \quad (33)$$

where $\lambda$= 365nm (GaN emission frequency), $t_0$ = 10f s, $t_p$ = 30f s.

For the coupled model, the expression for continuous dipole source is given by 8 and it requires modeling the time signal s(t) as discussed in section 2. Therefore, s(t) is modeled in such a way that its time response is continuous and the output spectrum consists of emission linewidth with Gaussian distribution. In this way, the coupling equation 8 that converts radiative recombination rates into continuous random dipole sources with emission frequencies forms the base of coupled model excitation source.

The continuous time signal with linewidth in the output spectrum s(t) is generated by summation of multiple continuous wave (CW) sources with random phases, where the weight of each CW source is given by a Gaussian distribution. Therefore, s(t) is given by the following equation

$$s(t) = \sum_{k=1}^{N} G(\lambda_k) \times \left(1 - \exp(-(t/t_p)^2)\right) \times \cos\left(\frac{2\pi}{\lambda_k}t - \phi_r\right) \quad (34)$$

$G(\lambda_k)$ is a Gaussian distribution function and $\phi_r$ is the random phase function.

Thus, the input time source for GaN LED coupled model simulation is a continuous and random current dipole source.

The optical properties of gold contacts are modeled using a six-pole Lorentz-Drude model [32] whereas the complex permittivity data of GaN in [33] and $Al_{0.3}Ga_{0.7}N$ in [34] used in GaN LED simulation is fitted using five-pole Lorentzian function as shown in figures 13 and 14. The EM field propagation inside GaN LED is described through magnetic field intensity distribution as shown in figures 15-17 for both optical and coupled model simulation. It can be seen from these figures that the light emitted in the active layer i.e. near the junction propagates through different layers inside LED that causes reflection, dispersion and power loss at different interfaces.

The emission spectrum of homojunction GaN LED is compared with the experimental result [35] as shown in Figure 19. The fabricated p-n homojunction GaN UV-LED [35] consisting of a p-type GaN layer doped at 2 x $10^{20}$cm $^3$, and an n-type GaN layer doped at 2 x $10^{17}$cm $^3$ emits at 370 nm as seen in Figure 19 whereas the simulated result has an emission around 366 nm with an emission linewidth of 20 nm. The dispersion effect can also be seen on EQE as shown in figure 20. The maximum EQE measured for this structure was 0.1755% at a forward voltage of 4.267 V. The experimental value of EQE for p-n homojunction GaN LED is 0.18% [18] at forward voltage of 4 V.

5. Simulation of 2D P-n-N Double Heterojunction GaN LED with subwavelength array of silver contacts

An application of the coupled model for investigating improved LED designs in both electronic and optical aspects is presented in this section. An AlGaN/ GaN DH LED with holes in the metal contact i.e. array of metal contacts is proposed to achieve enhanced optical power through extraordinary transmission (EOT) that occurs only for TM polarization.

Various studies have been proposed in the literature to increase the external efficiencies such as the use of metal nanoparticles, surface plasmons, nanoantennas, dielectric nanoantenna arrays, optical nanoholes [36, 37, 38] coupled with the light emission. The metal nanoparticles have the property to manipulate light at sub-wavelength scales and they can resonate at optical frequencies that can be tuned with their sizes, shapes and configuration. The basic idea in such studies is to resonate the metal nanoparticles at emission frequency of GaN LED structures. However, these investigations involve only optical simulation assuming that the enhancement is only due to the resonance of EM fields inside the device.

We therefore investigate AlGaN/GaN DH with array of silver contacts (length L, thickness H, spacing S, number of contacts N) as shown in figure 21 using the coupled model that can demonstrate the light enhancement resulting from both the electronic and optical parameters. The EM field enhancement can be achieved through EOT caused due to the localized surface plasmon resonances at the surface of metal contacts. Therefore, for the field enhancement, the array of metal contacts placed on top surface of GaN LED should resonate at emission wavelength.

This configuration is obtained by determining the scattered power spectrum solved using Total-Field Scattered-Field (TFSF) theory. At resonance, the contacts scatter more power than off resonance. Therefore, the peak of the scattered power spectrum should be con figured to position at GaN emission frequency. The configuration obtained for the resonance condition at 365 nm emission frequency is L=40 nm, H=50 nm, S=100 nm and N=2 to 12. It is further optimized by determining the transmission spectrum of this contact array configuration with AlGaN layers. The final configuration enhancing the optical power is therefore obtained for L=40 nm, H=50 nm, S=100 nm and N=10.

The coupled model simulation results for this structure are shown in figures 24-27. It can be seen from figures 24, 25 that the LED structure with silver contact array improves the current spreading in the device. This is attributed to the high potential distribution at the surface and also inside the GaN LED structure as compared to that of GaN LED with single p-contact. This effect can also be seen on radiative recombination rates shown in figure 26, where they are improved with the improvement in the current spread across the device. The improvements in the optical parameters can also be seen from figure 27, where they are enhanced with these contact configuration because of localized fields at the interface of AlGaN=Silver contact array.

Therefore, the combined effect of both electronic and optical model i.e. improved currents, radiative recombinations and enhanced EM fields is reflected in the coupled model output optical powers and EQE. With these improvements, the EQE of AlGaN/

GaN DH LED as compared to homo-junction GaN LED, increased from 0.1755 % to 1.825 % with the silver contact array.

6. Conclusions and Acknowledgments

An FDTD-DD algorithm for simulation of carrier transport and an ADE-FDTD algorithm for simulation of light propagation are presented. The major parameters involved in coupling the carrier and wave i.e. radiative recombination rates and optical generation rates are compared using coupled model, and it was found that the radiative recombination rates are more effective than the generation rates in coupling electronic and optical models. An appropriate relation describing the radiative recombination rates in terms of dipole sources is established. Therefore, an explicit time dependent dipole source representing continuous emission and whose strength is dependent on recombination rate is modeled for coupled model simulation.

The FDTD analysis of homojunction and heterojunction GaN LEDs is performed with the electronic, optical and coupled model. Double heterojunction GaN/ AlGaN LEDs with holes in the metal electrode (silver contact array) is investigated as an application of the coupled model, where EQE is improved by almost ten times as compared to that of homojunction GaN LED.

The coupled model can therefore be used to simulate real time domain carrier-wave interactions in III-Nitride LEDs that cannot be performed by steady state electronic simulators or FDTD optical simulators alone. There-fore, to simulate carrier transport in conjunction with light propagation with time, the time domain electronic model coupled with the optical model is expected to find innovative applications in the field of solid state lighting.

The authors wish to acknowledge the support of King Fahd University of Petroleum & Minerals in conducting this research work.

Table 1. Simulation Parameters [29, 23, 30, 31]

| Parameter | Value |
|---|---|
| Carrier concentration of n and p layers, $N_a$ and $N_d$ [cm$^{-3}$] | $10^{17}$ |
| Static Dielectric constant $\varepsilon_r$(GaN) | 9.7 |
| Band gap $E_g$(eV)(GaN) | 3.4 |
| Low field electron mobility $\mu_{no}$(cm$^2$V$^{-1}$s$^{-1}$) | 880 |
| Low field hole mobility $\mu_{po}$(cm$^2$V$^{-1}$s$^{-1}$) | 35 |
| Effective density of states (electrons) $N_c$(cm$^{-3}$) | $2.23 \times 10^{18}$ |
| Effective density of states (electrons) $N_v$(cm$^{-3}$) | $4.624 \times 10^{19}$ |
| Dielectric constant $\varepsilon_r$(Al$_2$O$_3$) | 3.218 |
| SRH recombination Coefficient A(s$^{-1}$) | $1 \times 10^5$ |
| Radiative recombination Coefficient B(cm$^3$s$^{-1}$) | $1 \times 10^{-8}$ |
| Auger recombination Coefficient C(cm$^6$s$^{-1}$) | $1 \times 10^{-30}$ |

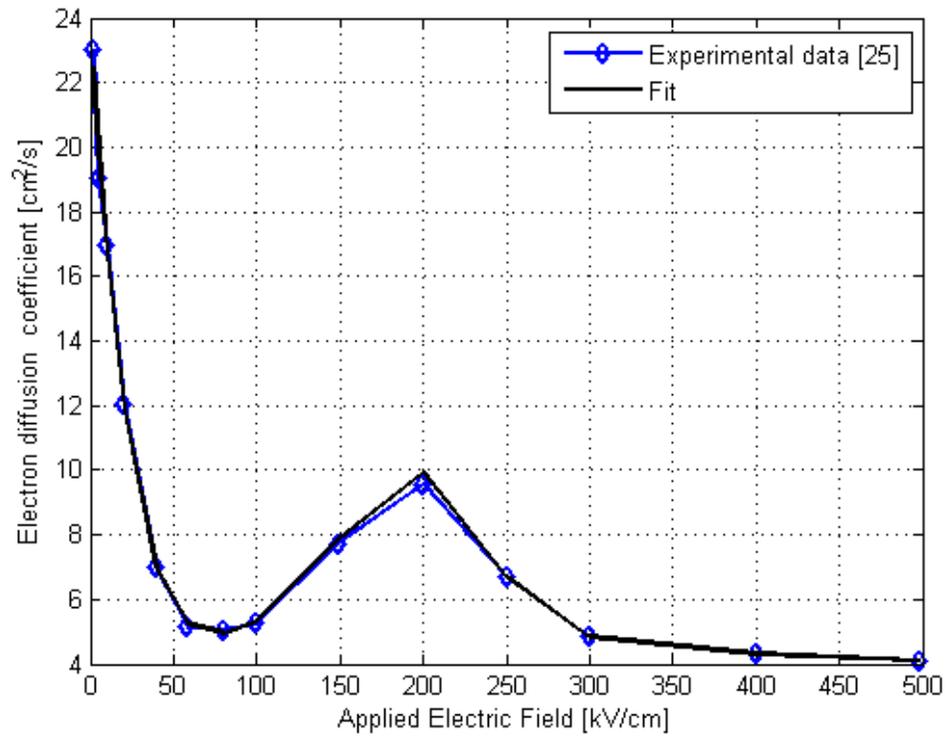

Figure 3. Electron Diffusion Coefficient of GaN vs. applied field.

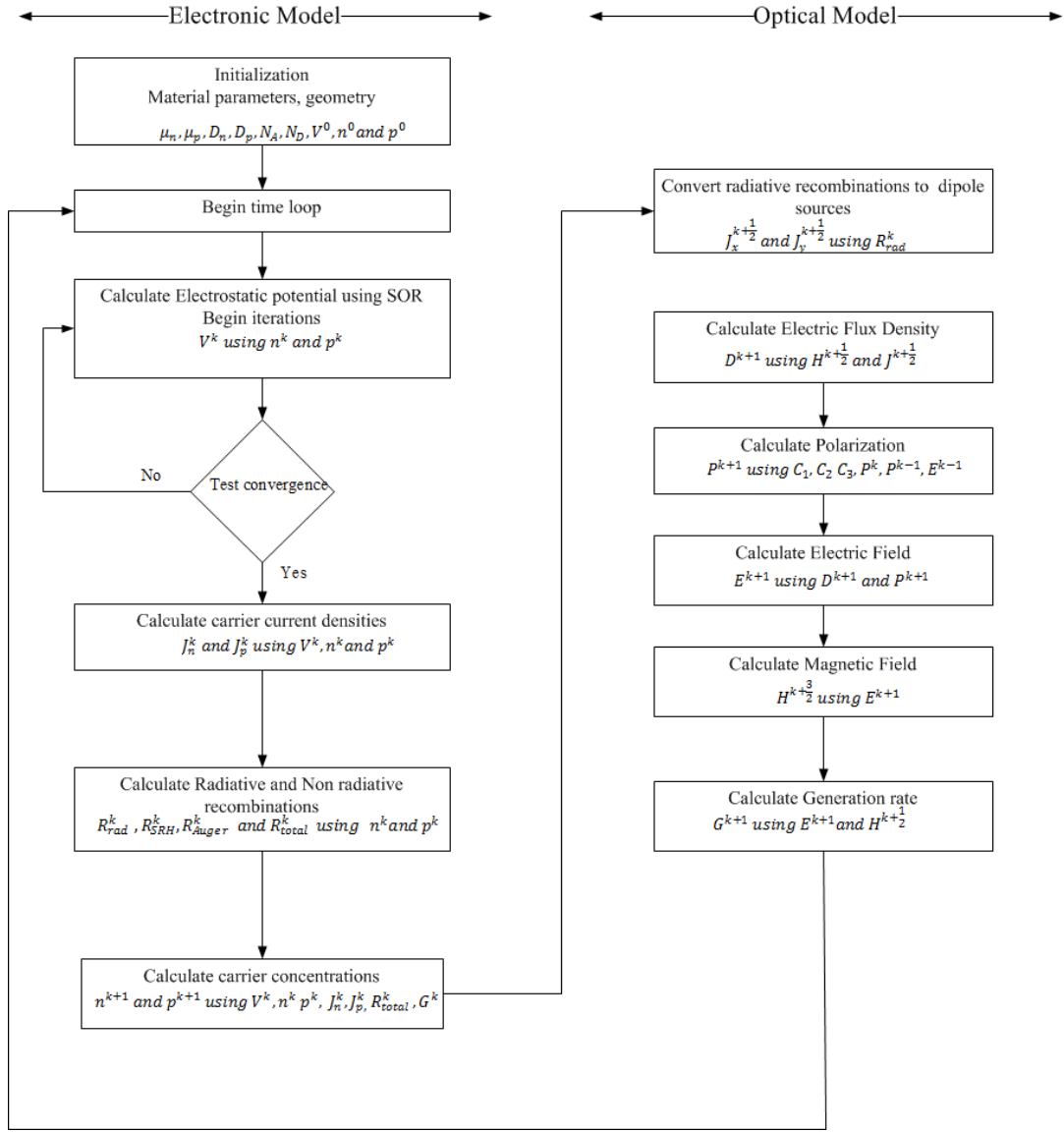

Figure 4: Coupled Model Algorithm

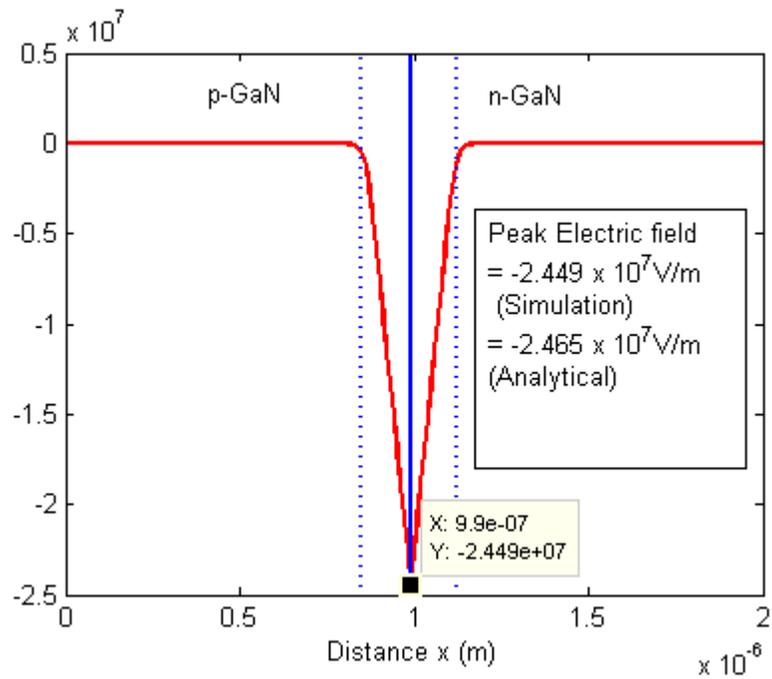
Figure 5. Simulated Electric field distribution at zero applied potential.

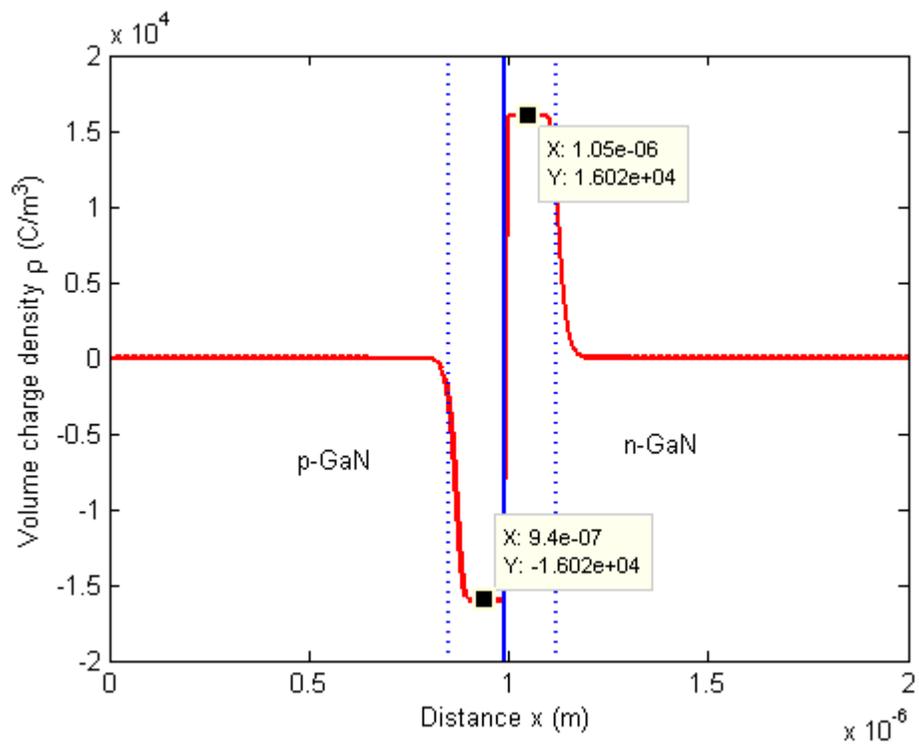
Figure 6. Simulated volume charge density at zero applied potential.

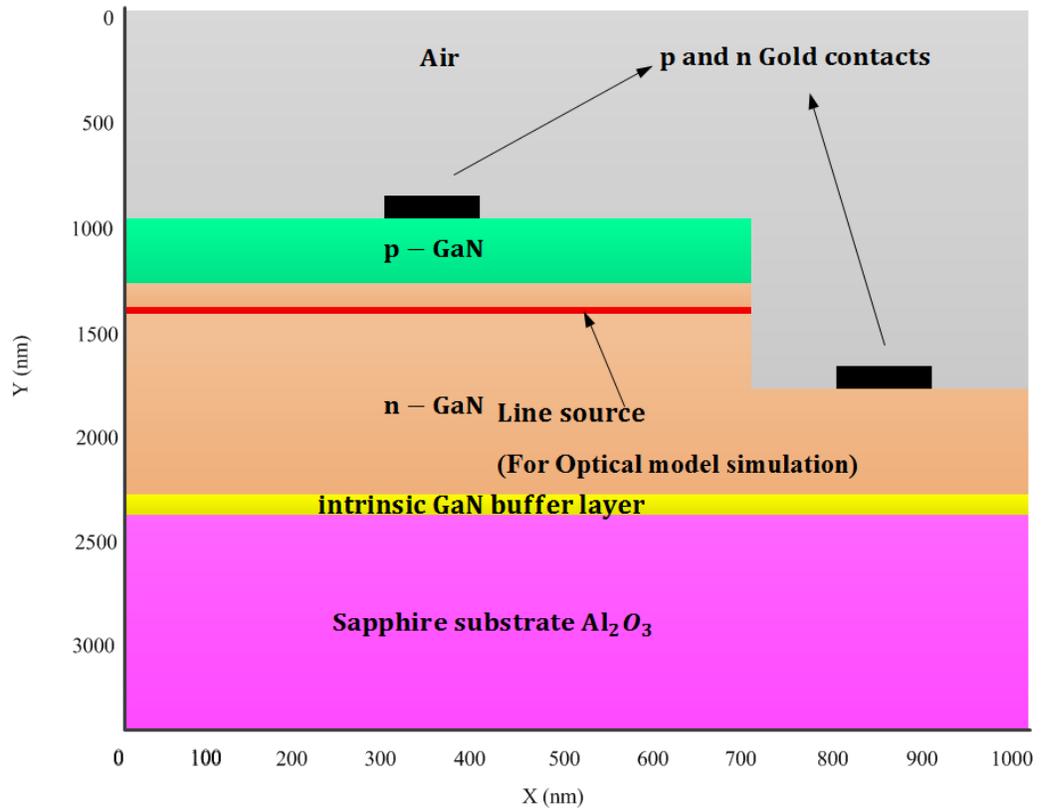

Figure 7. FDTD Simulation domain of p-n homojunction GaN LED.

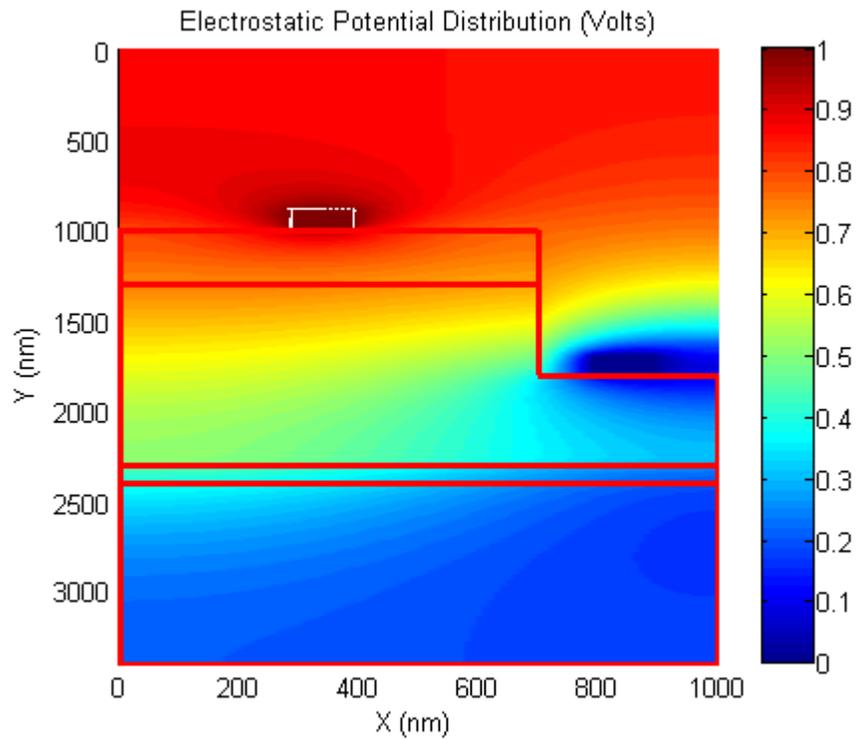

Figure 8. Simulated Electrostatic potential distribution of p-n homojunction GaN LED ($V_{applied} = 4.267V$).

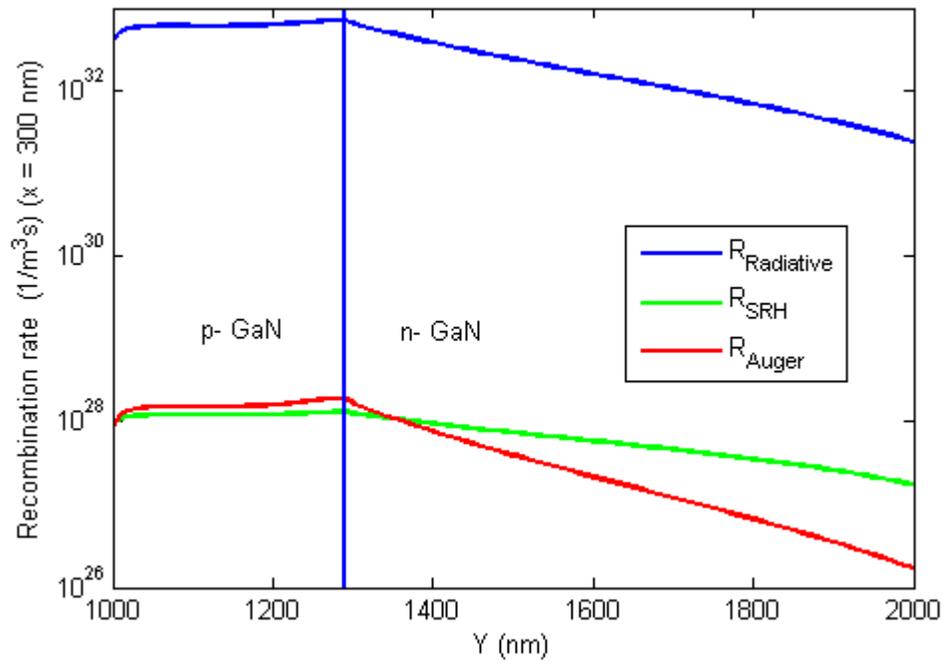

Figure 9. Electron hole pair recombination distribution- radiative and non-radiative ($V_{applied}$ = 4.267V ).

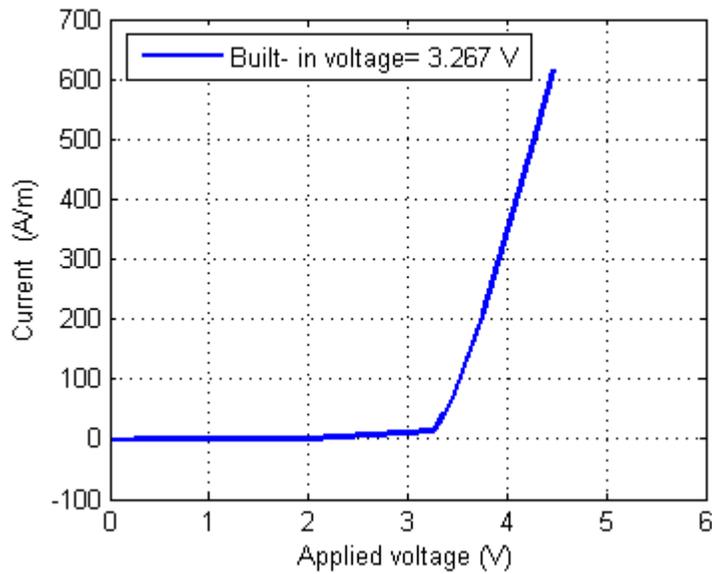

Figure 10. I-V characteristics of p-n homojunction GaN LED.

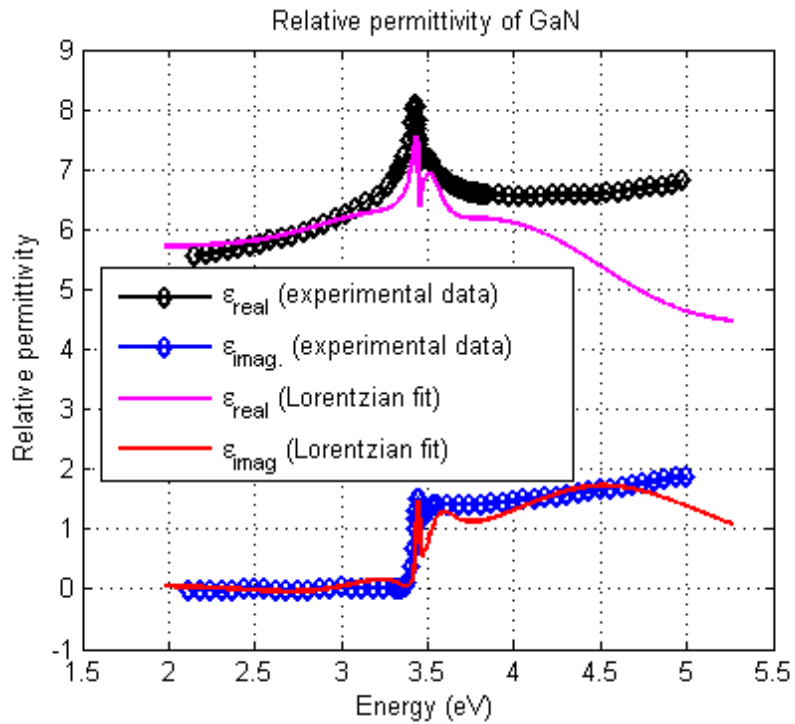

Figure 13. Relative permittivity of GaN ([33]).

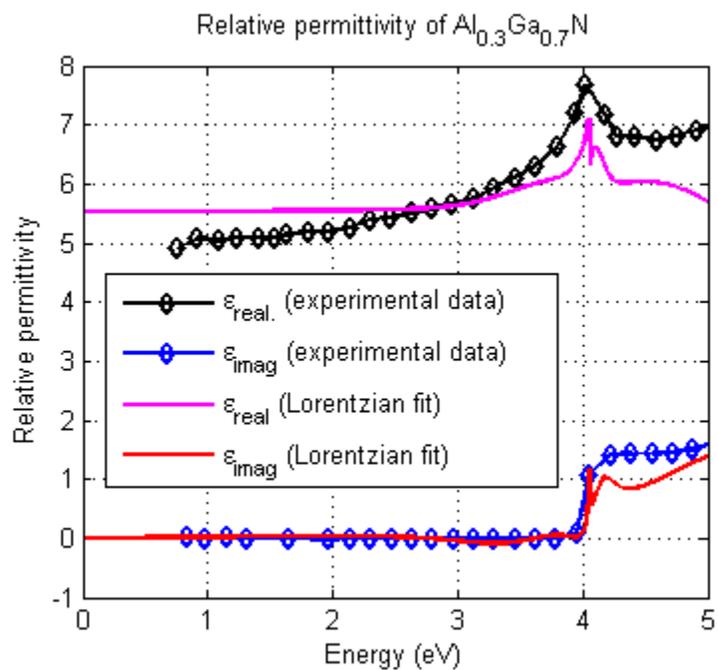

Figure 14. Relative permittivity of Al$_{0.3}$Ga$_{0.7}$N ([34]

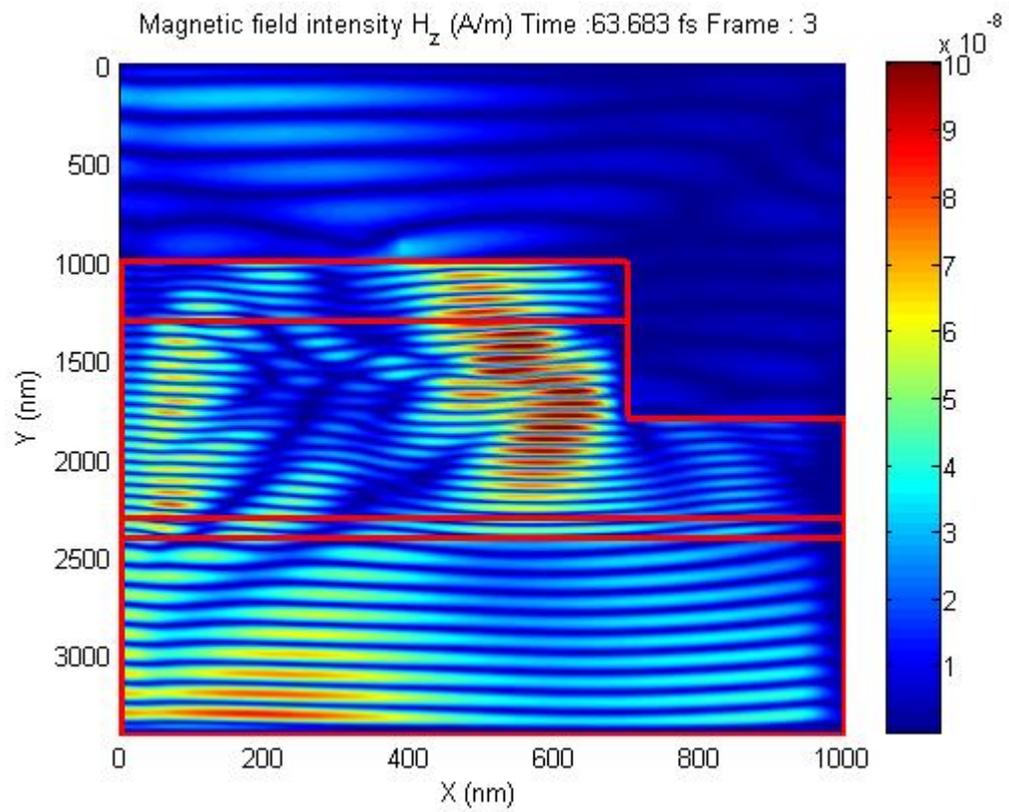

Figure 15. Snapshot of Magnetic field distribution-Optical Model.

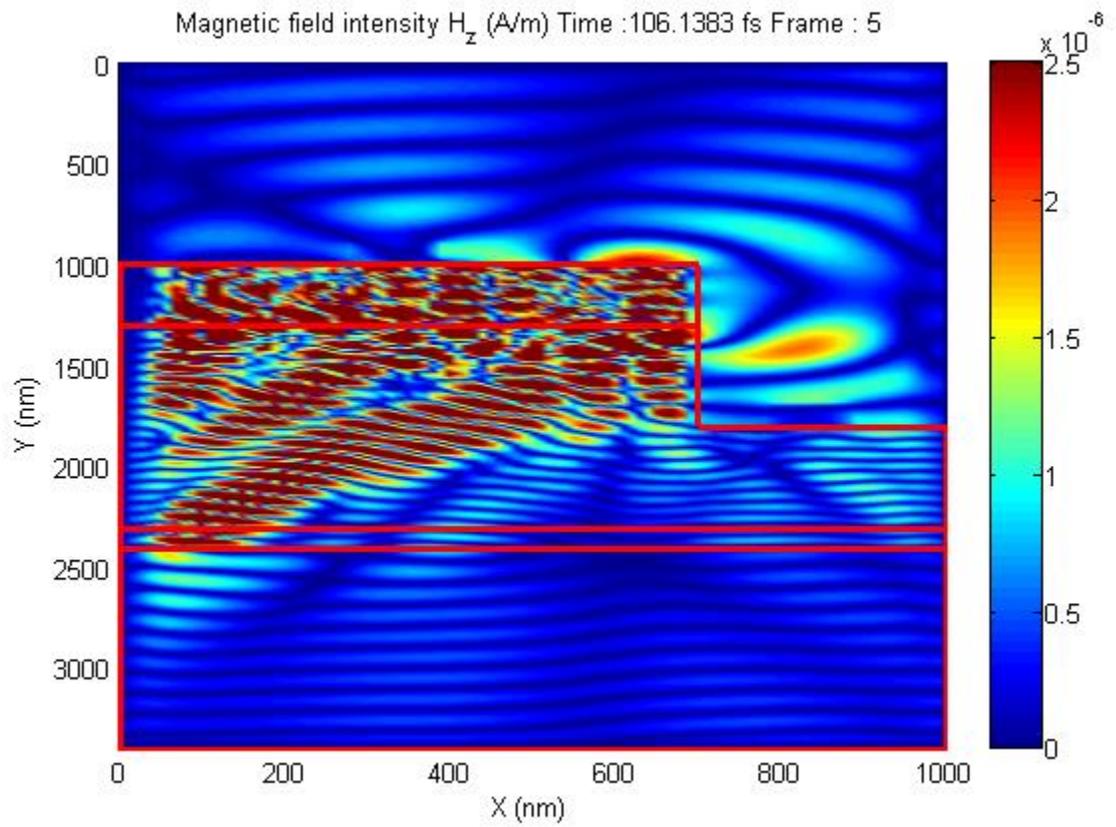

Figure 17. Snapshot of Magnetic field distribution-Coupled Model.

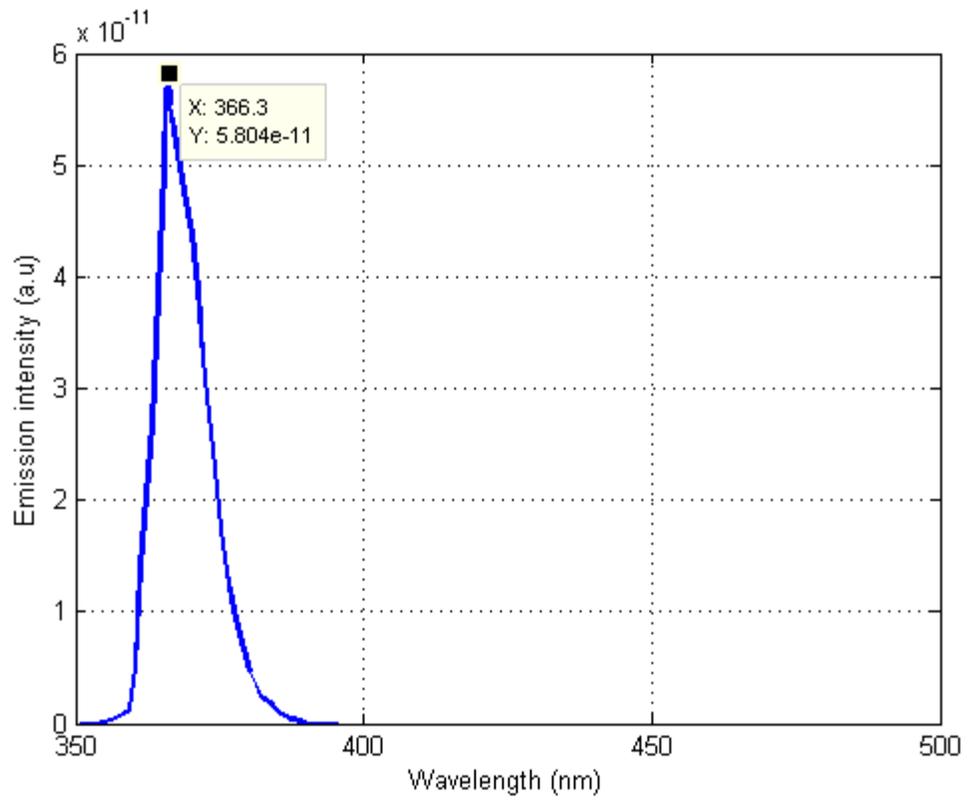

Figure 19. Emission spectrum of p-n homojunction GaN LED-Coupled Model.

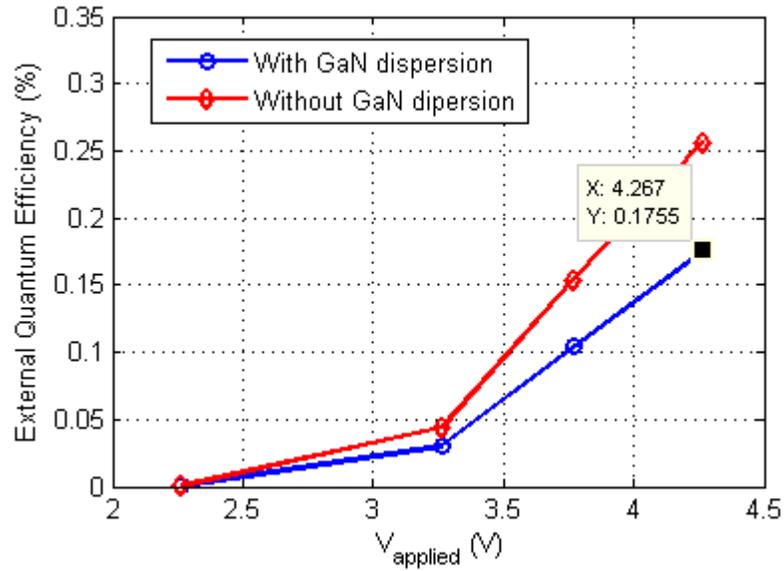

Figure 20. External quantum Efficiency (EQE) of p-n homojunction GaN LED vs. applied voltage.

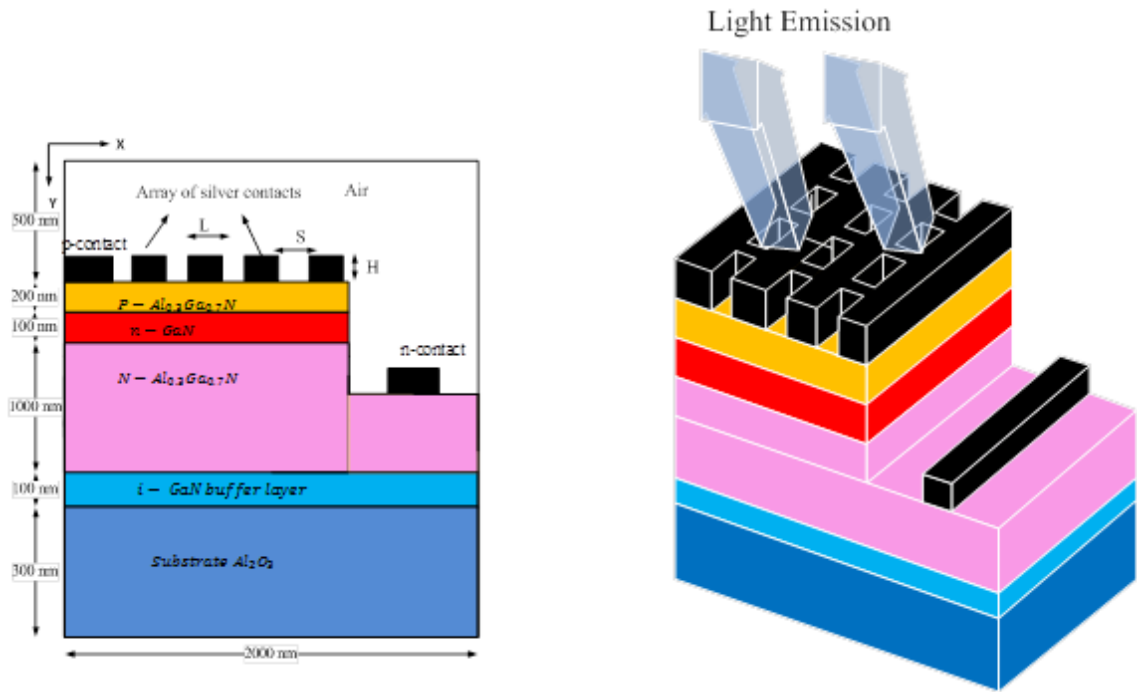

Figure 21. Schematic of AlGaN/ GaN DH LED with holes in the metal contact (silver contact array.

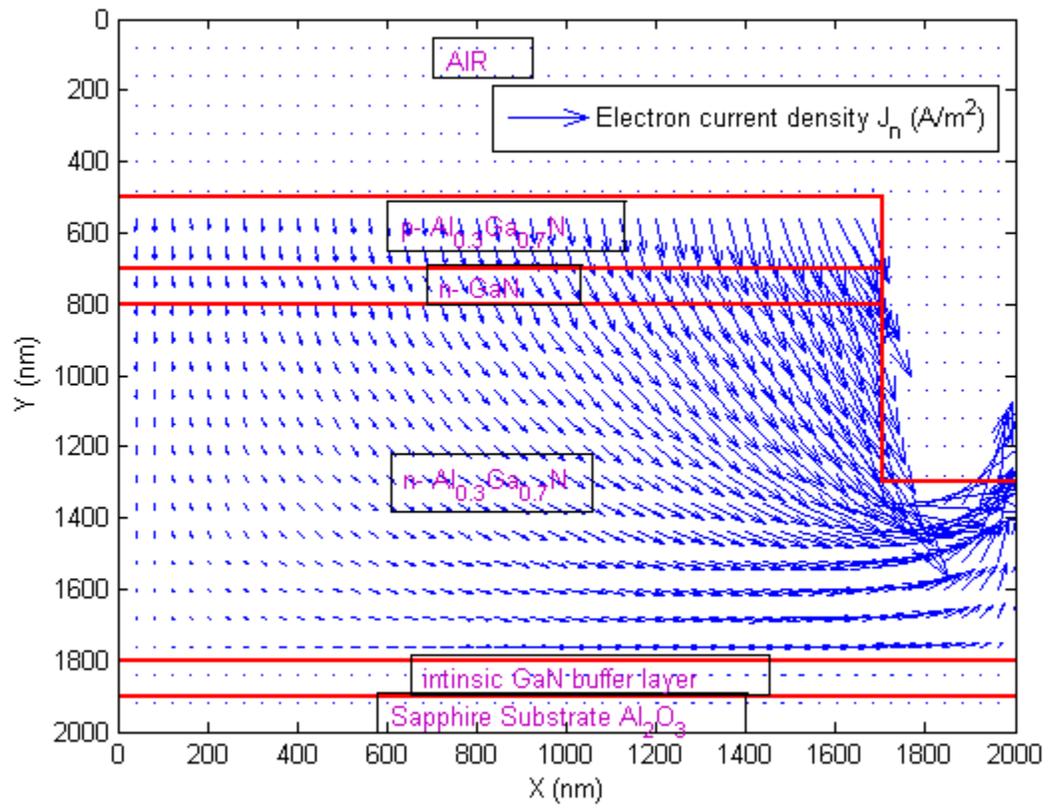

Figure 24. Electron current density distribution- with silver contact array

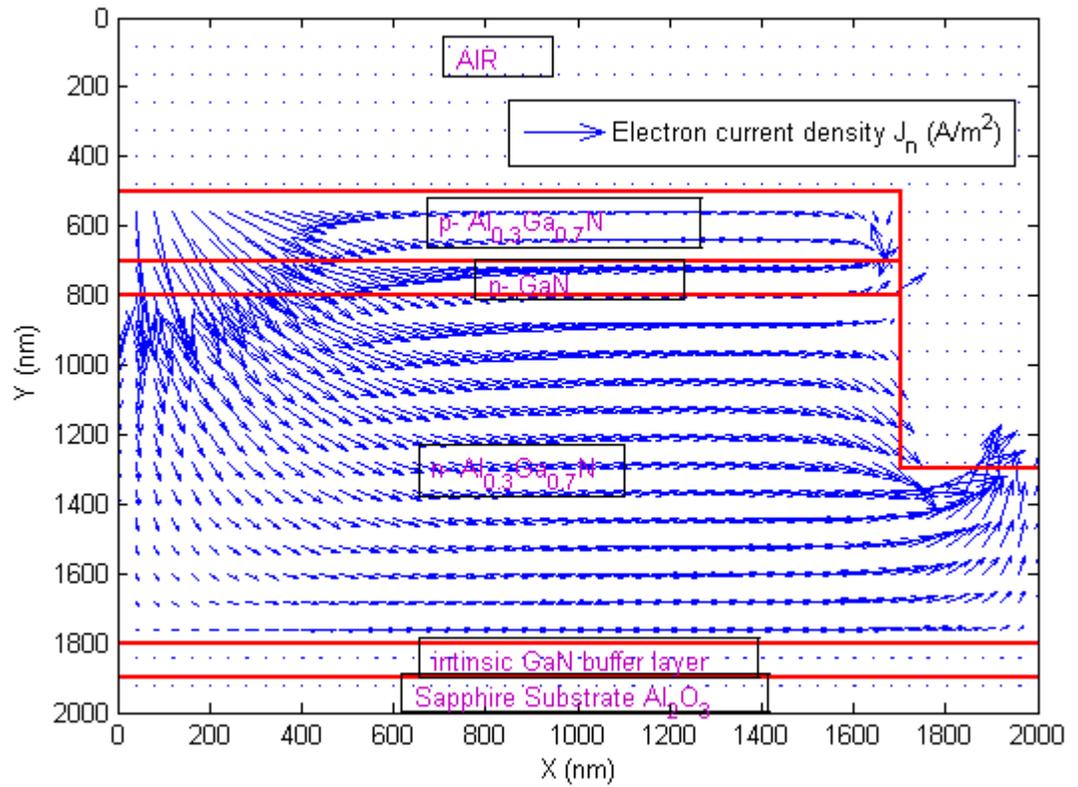

Figure 25. Electron current density distribution- without silver contact array.

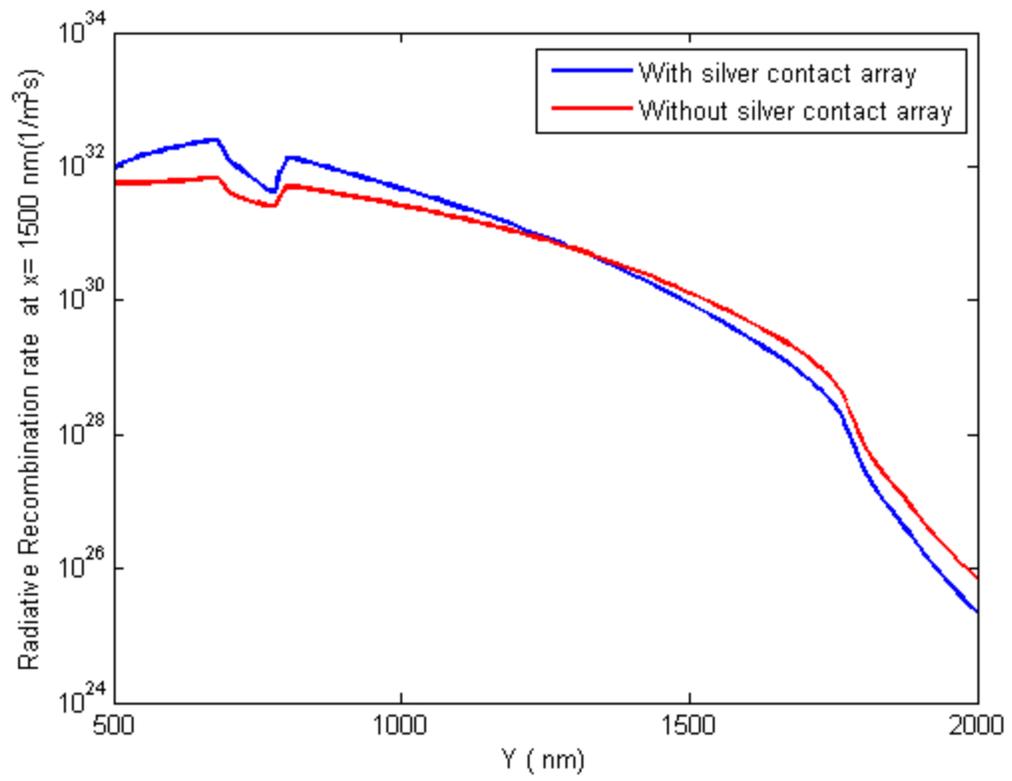

Figure 26. Radiative recombination rates- with and without silver contact array.

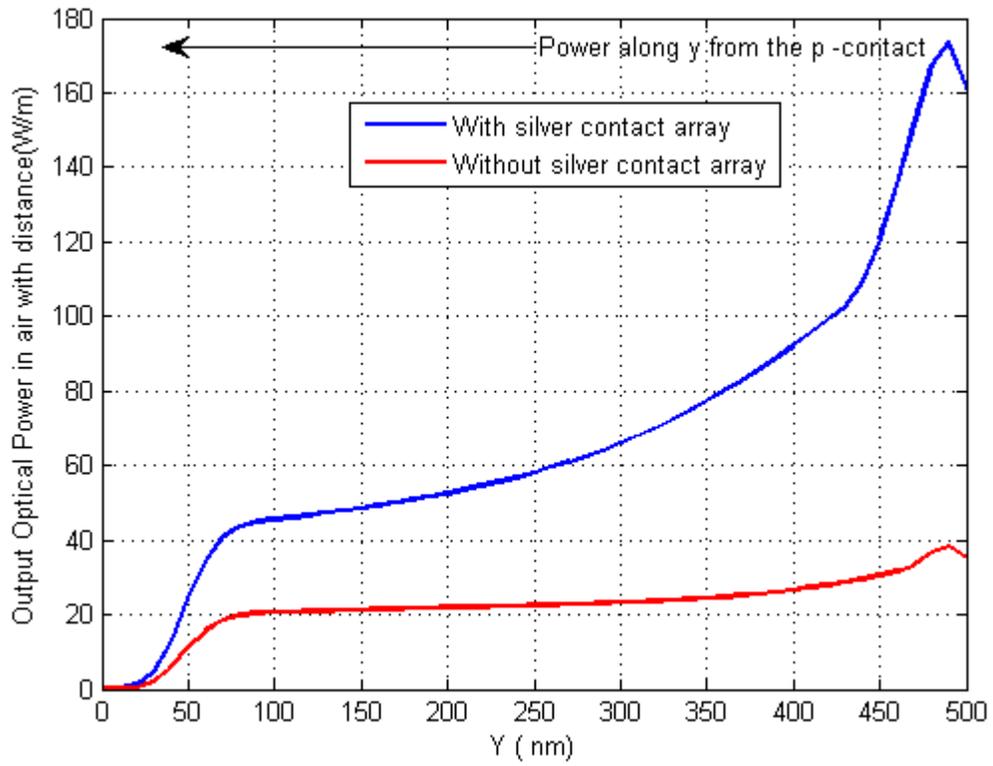

Figure 27. Output optical power from coupled model with and without silver contact array.